\documentclass[12pt,preprint]{aastex}

\slugcomment{Based on observations made at GEMINI-S, NOAO Proposal 
ID 2010B-0183.}

\shorttitle{Flourine in extragalactic carbon stars}
\shortauthors{Abia et al.}

\begin{document}

\title{The First Fluorine Abundance Determinations in Extragalactic AGB Carbon Stars}

\author{C. Abia\altaffilmark{1}}
\affil{Dpto. F\'\i sica T\'eorica y del Cosmos, Universidad de Granada, 18071 Granada, Spain}
\email{cabia@ugr.es}

\author{K. Cunha\altaffilmark{2}}
\affil{National Optical Astronomy Observatory, P.O. Box 26732, Tucson, AZ 85726, USA}

\author{S. Cristallo\altaffilmark{1,*}}
\affil{Dpto. F\'\i sica T\'eorica y del Cosmos, Universidad de Granada, 18071 Granada, Spain}
\affil{{\bf *Present address: INAF-Osservatorio di Collurania, 64100 Teramo, Italy} }

\author{P. de Laverny\altaffilmark{3}}
\affil{University of Nice-Sophia Antipolis, CNRS (UMR 6202), Cassiop\'ee, 
Observatoire de la C\^ote d'Azur, B.P. 4229, 06304 Nice Cedex 4, France}

\author{I. Dom\'\i nguez\altaffilmark{1}}
\affil{Dpto. F\'\i sica T\'eorica y del Cosmos, Universidad de Granada, 18071 Granada, Spain}

\author{A. Recio-Blanco\altaffilmark{3}}
\affil{University of Nice -Sophia Antipolis, CNRS (UMR 6202), Cassiop\' ee, 
Observatoire de la C\^ote d'Azur, B.P. 4229, 06304 Nice Cedex 4, France}

\author{V.V. Smith\altaffilmark{2}}
\affil{National Optical Astronomy Observatory, P.O. Box 26732, Tucson, AZ 85726, USA}

\and

\author{O. Straniero\altaffilmark{5}}
\affil{INAF-Osservatorio di Collurania, 64100 Teramo, Italy}

\begin{abstract}  Fluorine  ($^{19}$F)  abundances {\bf (or upper limits)} 
are derived  in  {\bf six} extragalactic  AGB carbon stars from the  HF(1-0) R9 line at
2.3358  $\mu$m in  high resolution  spectra. The  stars belong  to the
Local Group  galaxies LMC, SMC  and Carina dwarf spheroidal, spanning
more  than a  factor 50  in metallicity.  This is  the first  study to
probe  the  behaviour   of  F  with  metallicity  in
intrinsic extragalactic C-rich AGB stars.  Fluorine could be measured  only in four
of the  target stars, showing a wide range in F-enhancements.  
Our F abundance measurements together with those recently derived
in Galactic AGB carbon stars show a correlation with the observed carbon and $s-$element 
enhancements. The observed correlations however, display a different dependence on the
stellar metallicity with respect to theoretical predictions in low mass, 
low metallicity AGB models.  We briefly discuss the  possible reasons 
for this discrepancy.  If our findings are confirmed in a larger number of metal-poor  
AGBs, the issue  of F production in AGB stars will need to be revisited.

\end{abstract}


\section{Introduction} 
Comparisons between abundances of some  key  elements determined in
stars  in different  evolutionary stages  with
nucleosynthetic  stellar   models  is  a  fundamental   tool  for  our
understanding of stellar interiors. One element that  can add new
insight into this is fluorine, since this element is very sensitive to
nuclear  reactions involving proton  and/or alpha  captures. Actually,
the  origin of  this light  element is  still uncertain  although much
observational progress has  been made  in recent  years  in cool  K, M  and Asymptotic  
Giant Branch  (AGB) stars (e.g. Jorissen  et al.  1992, hereafter JSL;  
Cunha et  al. 2003-08; Smith et al.  2005; Uttenthaler et al. 2008),  and 
in post-AGB stars and planetary nebulae (Werner et al.  2005; 
Otsuka  et al. 2008). From  these observational  studies two  conclusions  
are derived:  i) AGB  stars
constitute  the only evidence  of stellar  F production
and, ii) the inferred evolution of the F abundance in the Galaxy, when
compared with chemical evolution  models (Renda et al. 2004), requires
the  contribution   of  the   additional  sources  proposed   so  far:
gravitational supernovae  (Woosley et  al. 1990) and  Wolf-Rayet stars
(Palacios et al. 2005).

Evidence  of F  production in  AGB  stars was  found by  JSL two
decades ago: F enhancements up to  a factor 30 solar and a correlation
of F with the  C/O ratio was first reported in Galactic (O-rich \&
C-rich) AGB  stars of solar  metallicity. Since the  C/O ratio is
expected to  increase in the  envelope during
the AGB phase as a  consequence of the third dredge-up (TDU) episodes,
this    was   interpreted    as   evidence    of   F
production. The F enhancements found by JSL, however, could
not  be accounted  by  detailed models  for  AGB stars  at that  epoch
(Mowlavi et al. 1998), or  by more recent ones
(Lugaro  et  al. 2004)  that  included  the impact  of the  variation of  
some nuclear  reactions with uncertain rates affecting F. Nonetheless, a 
recent re-analysis of the JSL's stars by Abia
et al. (2009-10; hereafter Paper I and II) using improved line lists
and model atmospheres found that the F  abundances reported in JSL have  
been overestimated because of  a  possible  lack  of  proper accounting  for  C-bearing  molecule
blends. This reanalysis reported F abundances by  $\sim 0.7$
dex lower on  average.  The  new F  abundances  and  the  observed
correlation  between  F and  $s-$element  enhancements  in solar metallicity C-stars, 
are now fully accounted by the current nucleosynthetic
modelling  of low mass AGB  stars (Cristallo  et  al. 2009-11). 

At   low metallicities the    situation   is   less
clear.  Theoretically, the  F production in  AGB  stars is
expected to increase when decreasing metallicity.  AGB stars synthesise  F  via
$^{14}$N$(n,p)^{14}$C~$(\alpha,\gamma)^{18}$O$(p,\alpha)^{15}$N$(\alpha,\gamma)^{19}$F and $   $
$^{14}$N$(\alpha,\gamma)^{18}$F$(\beta+)^{18}$O$(p,\alpha)^{15}$N$(\alpha,\gamma)^{19}$F, 
where the neutrons are provided by $^{13}$C$(\alpha,  n)^{16}$O and the protons mainly by
$^{14}$N$(n,  p)^{14}$C.  Thus,  the  production  of F 
basically  depends on  the  availability of  $^{13}$C  in the  He-rich
intershell, but
also on the amount of $^{13}$C available in the
ashes of  the H-burning shell. The  former is weakly  dependent {\bf on} the
stellar metallicity but the  latter scales  with the CNO
abundances in the  envelope, which, in turn, depends  on the (primary)
$^{12}$C dredged up during TDU episodes. As a consequence, the resulting
fluorine can be roughly considered  a primary element. Because of this
primary nature, larger  [F/Fe]\footnote{We  adopt  the usual notation with     
[x/y]$=$log    $[N(x)/N(y)]_{\star}-$     log
$[N(x)/N(y)]_{\sun}$    and    log    $\epsilon(x)=    12    +    $log
$[N(x)/N(H)]$.} ratios are obtained
in  metal-poor, low-mass  AGB   models  as  compared  with  the  solar
metallicity  case.  Nonetheless, the available determinations of F abundances in low
metallicity stars depict a more complex scenario than that expected from
this simple theoretical {\bf scheme}. For  instance,  Cunha  et
al. (2003) derived [F/O]$\la 0.0$ in RGB stars of
$\omega$  Cen {\it enriched} in  $s-$process  elements, which  is
difficult to  reconcile with metal-poor AGB stars  being significant F
contributors.  Recently, Lucatello  et  al.  (2011) found  lower
[F/Fe]  ratios than  theoretically expected  in a  sample  of Galactic
carbon enhanced metal-poor stars  enriched in s-elements (CEMP-s). The
implications  of  these findings  are  not  easy: first, 
the surface composition in the  $\omega$ Cen giants
is  hampered by  the  uncertain star  formation history and  chemical
evolution  of  this  cluster  and,  on  the  other  hand, the
abundances measured in CEMP-s stars (mostly binaries) {\bf might be affected
by the} dilution  of the  accreted
material onto the envelope of the secondary star. The amount of material
accreted  is  uncertain and, thus are {\bf in general} the
interpretation  of  the  abundances.  On  the  contrary,  F
abundances derived in intrinsic metal-poor AGB stars, i.e. stars which
own   their  chemical   peculiarities   to  internal   nucleosynthesis,
are free of these  problems\footnote{In Section 3 we show that the [F/Fe] ratio achieved
in the envelope in the C-rich AGB phase is almost independent on the initial
F content in the star.}. Unfortunately, the few metal poor Galactic AGB stars, 
as those observed in globular clusters, have masses too low to undergo TDU events 
and thus,
to  pollute the envelope  with the  nucleosynthesis  ashes of  the
He-rich  intershell. For  this reason,  the  AGB C-stars in  the
satellite galaxies  of the  Milky Way offer  a unique  opportunity to
study  the  $^{19}$F  production  at low  metallicity.  These  galaxies
are well  known to content metal-poor stellar populations.

Here  we  present for  the  first  time  F abundance  measurements  in
metal-poor AGB  carbon stars in stellar systems other  than the
Galaxy: the SMC, LMC and Carina dSph galaxies. These
C-stars are  significantly  more metal  poor  than the  Galactic
counterparts  so  far  analysed  providing  valuable information on  how  F  is
produced in AGB stars as a function of metallicity.

\section{Observations  and  Analysis}  
The  stars  were  chosen from  previous optical  high-resolution  spectroscopic
studies of  extragalactic AGB  C-stars (de  Laverny et  al. 2006;
Abia et al. 2008). In addition to the stars analysed there (BMB
B30 belonging to the SMC, and ALW-C6 and ALW-C7 to Carina dSph), we 
added two stars  in the LMC (TRM88  and MSX663) and a new target in the
SMC (GM780). Details about the  characteristics of BMB B30, ALW-C6 and
ALW-C7 can be  found in the above works (the other stars are described below).
The stars were  observed in classical mode with  the 8.1 m Gemini-South
telescope  and the  Phoenix spectrograph  (Hinkle  et al.  1998) at  a
resolving power R$\sim 50,000$; and centred at 23350 {\AA},to  include the HF(1-0) R9  line. In  
Paper I it was shown that 
this line  is the most reliable for  F  abundance  determinations  in  cool  stars.  A  detailed
description  of the  Phoenix observations  and the  corresponding data
reduction  can be  found  in Smith  et  al.  (2002).  Table 1  
shows the exposure  times for each
object, and the signal-to-noise ratios  reached in the final spectrum.

Two target stars are peculiar: MSX663 is  a long-period
variable classified as a S star by Cioni et al. (2001)  
(C-rich but with  C/O$<1$).   Zijlstra  et  al.  (2006)
indicated that this object may be a symbiotic star.  Surprisingly, our
spectrum of this object shows no spectral lines in the 2.3 $\mu$m region; 
even the vibration-rotation CO lines, usually strong, 
are absent.  This might be  compatible with  this object being  a supergiant rather
than an AGB star.  In any case, we could not identify  the R9 line and
thus, it  was discarded for analysis. GM780 is
also a rare  object. Lagadec et al.  (2007) noted that  
the C$_2$H$_2$ bands, commonly  observed in
C-stars, were  absent.  Also,  the $J-K$  colour of  this object
(2.6), is quite red, and dust  should be present around it. These authors
concluded that  GM780 has a C/O  ratio considerably lower than
that in any other star in their sample.  Our high-resolution spectrum
confirms  this  finding; all  the  features  of CN and C$_2$
molecules  in the 2.3  $\mu$m region appear very
weak.  Also, the high excitation CO lines look broader and affected  by line 
doubling. Actually, a much larger macroturbulence value than 
typical one was required to fit the line profiles in this star. We do not 
know the reason of this, but  the effect  of dust or/and stellar  pulsation might 
be  at play (see e.g.  Nowotny  et al.  2011). In
fact, our synthetic fit to its spectrum was not satisfactory.

The classical  method  of spectral  synthesis  was used  in the
analysis.   Theoretical  LTE spectra  were  computed in spherical geometry 
and convolved  with Gaussian functions to mimic the corresponding instrumental 
profile  adding a macroturbulence velocity typically of  $5-7$ km/s 
(a 10 km/s value was used for GM780). For more details on the method of analysis, 
and the adopted molecular and atomic line lists used see Paper I.
We  adopted  the  atmospheric parameters  derived  in  de  Laverny  et
al. (2006) and  Abia et al.  (2008) for the stars  BMB B30, ALW-C6 and
ALW-C7, and followed the same procedure to derive the stellar
parameters in the remaining stars (see  these works  for details). Accordingly,
we estimate  a T$_{eff}\sim 2000$ K for GM780 (note the red $J-K$ colour of this star),
and 3400 K for TRM88. Since our grid of C-rich atmosphere models (Gustafsson et  al. 2008) 
does not include a T$_{eff}$ as low as 2000 K, we used a model with such a T$_{eff}$ from 
the C-rich models grid by Pavlenko \& Yakovina (2010). 
A gravity of  log g$=0.0$ and a $\xi=2.2$ km/s were adopted 
for all the stars (see e.g. Lambert et al. 1986). The analysis of
BMB B30, ALW-C6 and ALW-C7 resulted in different C/O ratios 
than those previously derived from optical spectra. 
These differences were, nevertheless, within the uncertainties\footnote{The derived C/O 
ratios have an additional uncertainty since the O abundance cannot be determined 
independently of the C abundance. This is because theoretical spectra are almost 
insensitive to a large variation of the O abundance provided that the difference 
log $\epsilon($C$-$O) is kept constant. {\bf The estimated error in the C and O abundances
is $\pm 0.3$ dex. See de Laverny et al. (2006) and Paper I for details.}}.
Unfortunately, the observed   spectral   range   does   not   contain   useful   metallic
lines to estimate the metallicity ([Fe/H]) in the stars TRM88 and GM780. Thus, we initially adopted  
the metallicity of  the main stellar  population in
the LMC and SMC ([Fe/H]$\sim  -0.4$   and  $-0.7$,
respectively). Then, comparisons of synthetic to observed spectra
provided new estimates of the  metallicity by assuming that the O abundance derived
from  CO lines is an indication of the metallicity. This
procedure  was repeated  until a  good fit  to the  full  spectrum was
obtained. Despite the O abundance might not be a good indicator of
the metallicity, note that variations  
in the  metallicity of  the model atmosphere scale linearly to the F abundance derived.  
This means that the [F/Fe] ratio is almost independent of the metallicity  
adopted in  the model  atmosphere.  

Figure 1 shows examples of synthetic  fits in the R9 line region for
the  stars  TRM88 and  ALW-C7. The  blend  to the left
wing of the HF line has   a  significant   contribution   of
$^{12}$C$^{17}$O. This feature allowed  to derive the $^{16}$O/$^{17}$O  ratio 
(see Table 1) in  the most  metallic stars of  the sample. In the most metal-poor  
objects (stars in Carina) however,  this blend  is insensitive  to
variations of  the $^{16}$O/$^{17}$O ratio.  For some  of the targets,
Na  abundances  were  also derived  from  the  Na I line at $\lambda 23379$  {\AA}. 
This line is, however, blended with molecular features, thus its detection depends 
on the current Na abundance in
the star (see Table 1). For each element (and the $^{16}$O/$^{17}$O  ratio),
the uncertainty was calculated by determining individually  the sensitivities
of the derived abundance to the adopted T$_{eff}$, gravity, C and O abundances, 
microturbulence, and metallicity. We then sum in quadrature the resulting uncertainties 
associated to each parameter. The synthetic fit to the HF and the
Na I lines is particularly sensitive to the T$_{eff}$ adopted, the other
parameters affecting at a lower degree {\bf (in particular gravity and metallicity, see Paper I)}. 
The resulting formal uncertainty is $\pm 0.30$ dex for F, $\pm 0.26$ dex for Na, and a factor $\sim 2$ for the 
$^{16}$O/$^{17}$O ratio.

\section{Results and Discussion}

Table 1 summarises the main results. We find
a wide range in [F/Fe] ratios in metal-poor extragalactic AGB C-stars. Fluorine
is enhanced in all except one star, confirming that this element is produced during 
the AGB phase. Figure 2 provides further evidence of this. The similarity in the run of 
F with C strongly support that the synthesis of F is tied to the
production of C during the Thermal Pulsing (TP) AGB phase. Also Figure 2 qualitatively 
shows that F enhancement increases for decreasing stellar metallicity 
(although the star B30 seems to deviate from this trend). In 
Paper II we showed that at solar metallicity, the F and C correlation can be reproduced 
by current TP-AGB models (Cristallo et al. 2009, continuous lines in Figure 2). 
Models and observations, however, now seem to disagree in metal-poor AGB stars 
(dashed lines); theoretical models tend to overproduce C for a given F 
enhancement. Indeed, the two stars in Carina 
should be along the model computed with $Z\sim 10^{-4}$ (right dashed line), 
while the star B30 along the model line with $Z\sim 10^{-3}$. 
This discrepancy, which has been already noted at solar Z (e.g.
Abia et al. 2002), becomes more evident at lower metallicity as Figure 2 shows, 
since in current AGB models the efficiency of the TDU increases at low metallicity 
and thus, the amount of C dredged-up into the envelope. Note that this problem 
could be also an observational bias: intrinsic AGB C-stars with high C/O ratios should 
exist, as high values of C/O have been observed both in post-AGB stars and in extrinsic 
C-rich objects originated by mass transfer from an AGB star. However, in cool C-stars 
an excess of carbon is immediately translated into a copious production of C-rich dust 
and into a high opacity of the circumstellar environment, to the point of hiding 
the photosphere at visual wavelengths.
Most of the C might be trapped into grains. 
These dusty C-stars should show large infrared excess. Curiously enough, the two stars in 
our sample (Table 1) with high C/O ratios do not show this as deduced from 
their available infrared photometry, while the star GM780, with large infrared 
excess (see Section 2) has the typical C/O$\ga 1$ ratio observed in C-stars.

Figure 3 provides another piece of information on the synthesis of F in AGB stars. It
shows the observed relation between F and the average\footnote{Obtained as the mean value
of the [Sr, Zr, Y, Ba, La, Nd, Sm/Fe] ratios.} $s$-element enhancements in C-stars of
different metallicities compared with theoretical predictions. Again the observed trend 
{\bf points-out} to a correlation between [F/Fe] and [$<$s$>$/Fe]. Despite the scarce number of
metal-poor objects studied, the observations suggest a different dependence than that
theoretically expected (Cristallo et al. 2011) as a function of the metallicity: 
in metal-poor C-stars larger [F/Fe] ratios are expected for the observed 
$s$-element enhancement. While the F and Na enhancements\footnote{Na can be also produced
during the TP-AGB phase, see Cristallo et al. (2009).} found in ALW-C7 can be 
simultaneously accounted (within error bars) by our reference 
1.5 M$_\odot$, $Z\sim 10^{-4}$ TP-AGB model, the 
predicted [$<$s$>$/Fe] ratio is lower than the observed value. 
It is a challenge that the same models which nicely reproduce the trend found at solar 
metallicity 
(lower dashed line in Figure 3), fail in reproducing the metal-poor data. 
This discrepancy can also be noticed in Figure 4, where the [F/$<$s$>$] ratios found in 
Galactic CEMP-s (Lucatello et al. 2011) and intrinsic AGB C-stars are plotted. Note 
that by plotting the [F/$<$s$>$] 
ratio we avoid any dilution problem that may affect the abundances derived in CEMP-s 
stars. From this figure, it is clear again that the F content derived in metal-poor stars 
is lower than the predicted values. 

Now the question is how to account for these apparent low F enhancements?
In the recent years, some degree of extramixing processes have been invoked to explain
the $^{16}$O/$^{17}$O/$^{18}$O ratios found in grains of AGB 
origin, and the low $^{12}$C/$^{13}$C ratios in C-stars (Busso et al. 2010). 
Extramixing has been proposed by Denissenkov et al. (2006) 
{\bf as a possible solution}{\footnote{{\bf There are other explanations more widely accepted, see e.g. 
Marino et al. (2008).}}} of the observed anti-correlation of F and Na abundances 
found in low-mass giants of M4 (Smith et al. 2005). Extramixing 
might reduce F in the AGB envelope if such processes expose material at temperatures
high enough to activate the $^{19}$F$(p,\alpha)$ reaction. However, 
the $^{16}$O/$^{17}$O ratios found in TRM88 and B30 are those characteristic 
of stars which have undergone the first dredge-up. The measurement of the 
$^{16}$O/$^{18}$O ratio in these stars would be critical to give a definite answer
on the ocurrence or not of extramixing processes but, unfortunately, $^{18}$O could
not be determined. Even the large ($>100$) $^{12}$C/$^{13}$C ratio found in B30 
(de Laverny et al. 2006) cannot add any hint on the presence of 
extramixing, due to the larger $^{12}$C/$^{13}$C ratios attained on the surface 
according to metal-poor AGB models already after the first TDUs. On the other
hand, several nuclear reaction rates affecting F are still uncertain, in particular 
$^{15}$N$(\alpha,\gamma)^{12}$C. We performed {\bf preliminary} tests with nuclear rates
modified within the current uncertainties and found that this can only mildly improve the
situation but actually it seems not enough to solve the problem. Another possibility
is that these metal-poor C-stars formed with an initial F abundance much lower than the
one scaled to their metallicity (see Section 1). Note that the chemical evolution of these satellite
galaxies is essentially unknown (even the run of F with [Fe/H] in our Galaxy is 
uncertain). Nonetheless, according to our theoretical metal-poor AGB models 
the F production is so large that the [F/Fe] ratio reached in the envelope in the
C-rich phase (C/O$>1$) is independent of the initial F content in the star. For instance,
in a 1.5 M$_\odot$, Z$\sim 10^{-3}$ AGB model assuming an initial [F/Fe]$=-1.0$, the predicted
[F/$<$s$>$] ratio after a few TPs {\bf differs less than 0.1 dex with respect to that obtained 
with a solar scaled initial ratio. The difference is even smaller for lower metallicity models.} 

In summary, from the comparison between the derived F abundances and the 
current models we conclude that the F synthesis in metal-poor AGB stars is probably 
not {\bf as large as expected or some physical mechanism, not currently considered in the
models, efficiently destroys it}. Obviously, additional 
F abundance measurements in metal-poor AGB stars would be extremely important to enlighten 
this problem. The origin of this element still remains unknown.
  
\acknowledgments
Part   of   this   work   was   supported  by   the   Spanish   grants
AYA2008-04211-C02-02 and  FPA2008-03908 from  the MEC.  P.  de Laverny
and  A. Recio-Blanco  acknowledge the  financial support  of Programme
National  de Physique Stellaire  (PNPS) of  CNRS/INSU, France. S.C and O.S. 
have been partially supported by the italian
MIUR grant FIRB-Futuro in ricerca 2008. We are
thankful  to B.  Plez for  providing us  molecular line  lists  in the
observed infrared domain and to K. Eriksson and Y. Pavlenko for 
the metal-poor C-rich atmosphere models.

{\it Facilities:} \facility{GEMINI-S (Phoenix)}.

\clearpage

\begin{figure}
\epsscale{1.0}
\plotone{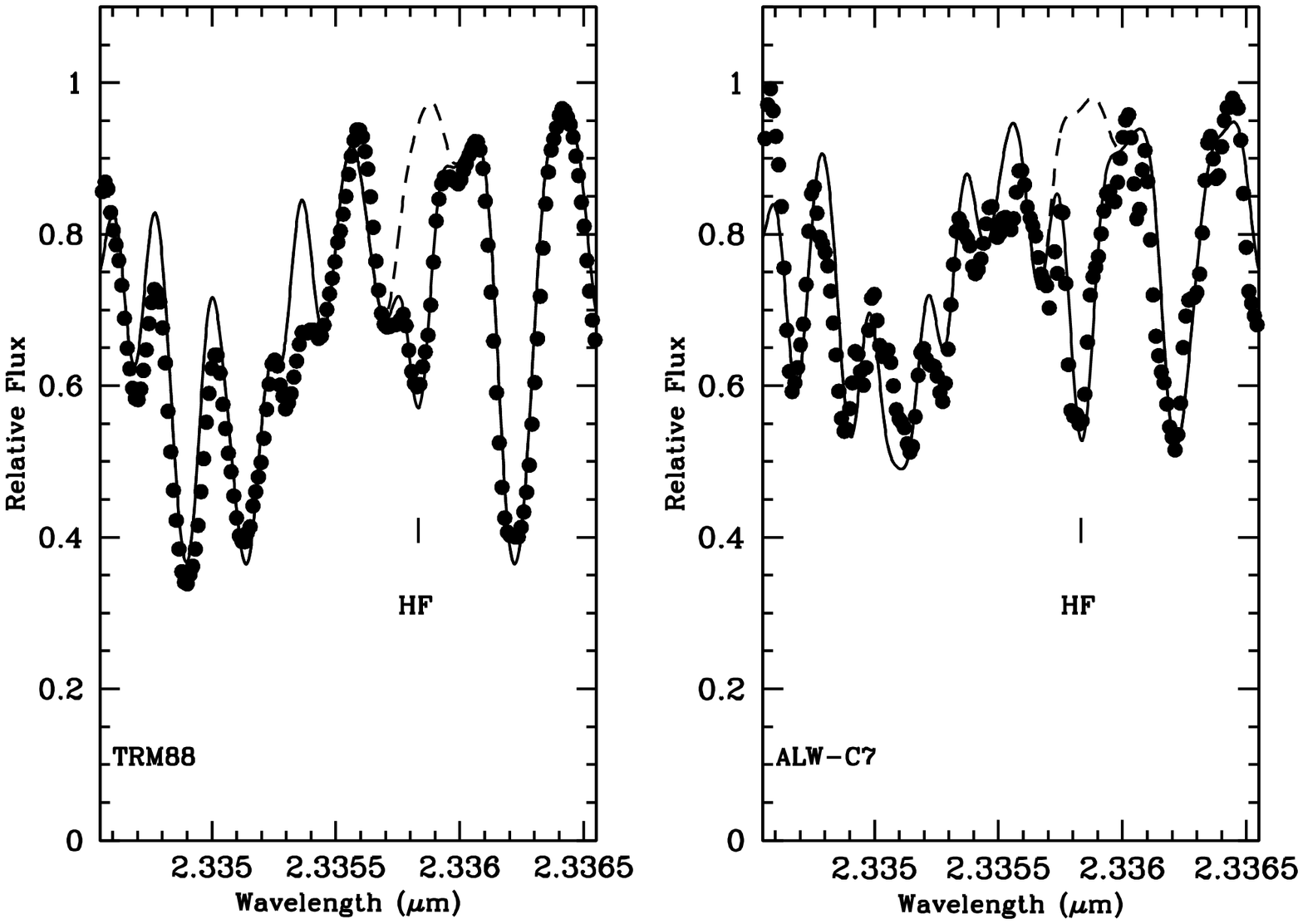}
\caption{Observed spectra in the region of the HF R9 line of the stars 
TRM88 and ALW-C7 (black dots) compared to their best fit model 
spectra (continuous lines) computed with the corresponding F abundances (see Table 1).
Synthetic spectra computed with no F (dashed lines) are also shown. Note the more
noisy spectrum of ALW-C7.\label{fig1}}
\end{figure}

\clearpage
\begin{figure}
\epsscale{1.0}
\plotone{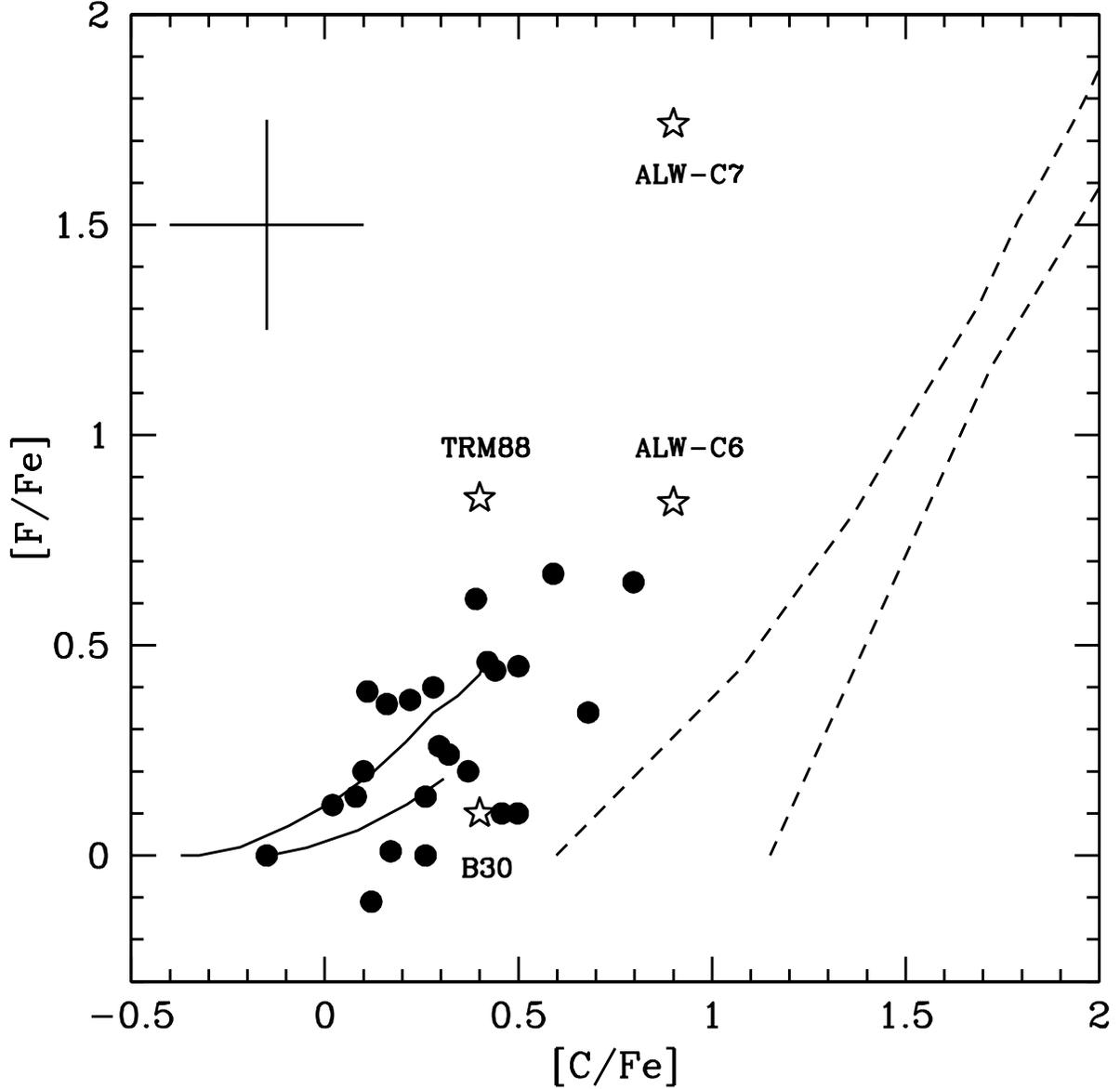}
\caption{Observed [F/Fe] vs. [C/Fe] ratios in Galactic 
(dots, from Paper II) and extragalactic (open stars, this study) N-type 
AGB carbon stars. Lines are theoretical predictions for a 2 M$_\odot$ and 1.5 
M$_\odot$ TP-AGB model with $Z=Z_\odot$ (left and right continuous lines, 
respectively) and for a 1.5 M$_\odot$  model with $Z=10^{-3}$ and $10^{-4}$ 
(left and right dashed lines, respectively). Note that for the 
[F/Fe] ratios derived in the metal-poor extragalactic carbon stars the predicted 
C enhancement is much larger than observed (see text).\label{fig2}}
\end{figure}
\clearpage
\begin{figure}
\epsscale{1.0}
\plotone{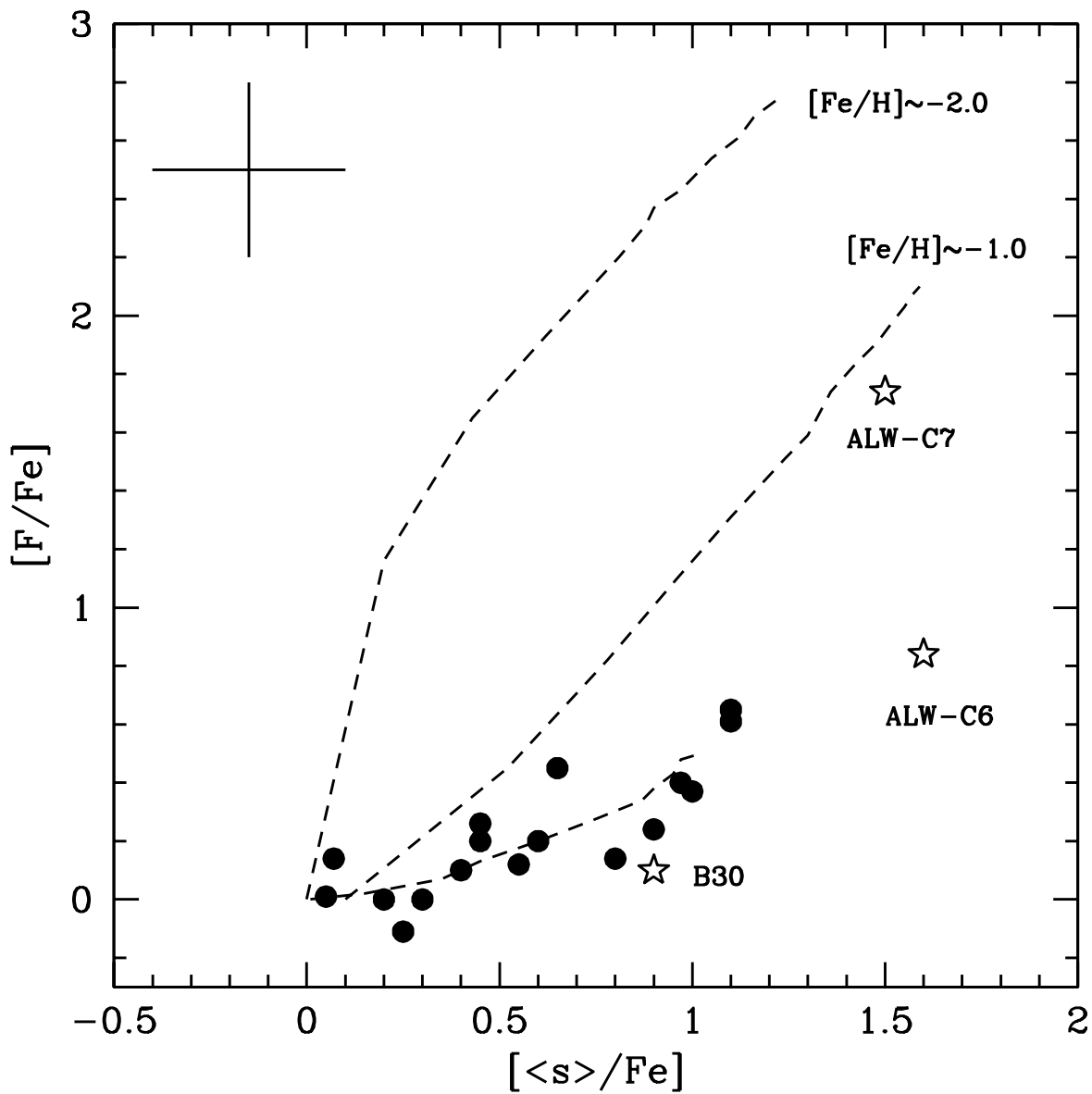}
\caption{Fluorine vs. average $s$-element enhancements in Galactic and
extragalactic AGB carbon stars. Symbols as in Figure 2. Dashed lines
are theoretical predictions for a 1.5 M$_\odot$ TP-AGB model for different
metallicities ([Fe/H]$\sim 0.0, -1.0, -2.0$ from bottom to up) according
to Cristallo et al. (2011) (see text). The [$<s>$/Fe] ratios in the extragalactic
stars are from de Laverny et al. (2006) and Abia et al. (2008), while those
in Galactic stars from Abia et al. (2002).\label{fig3}}
\end{figure}

\clearpage
\begin{figure}
\epsscale{1.0}
\plotone{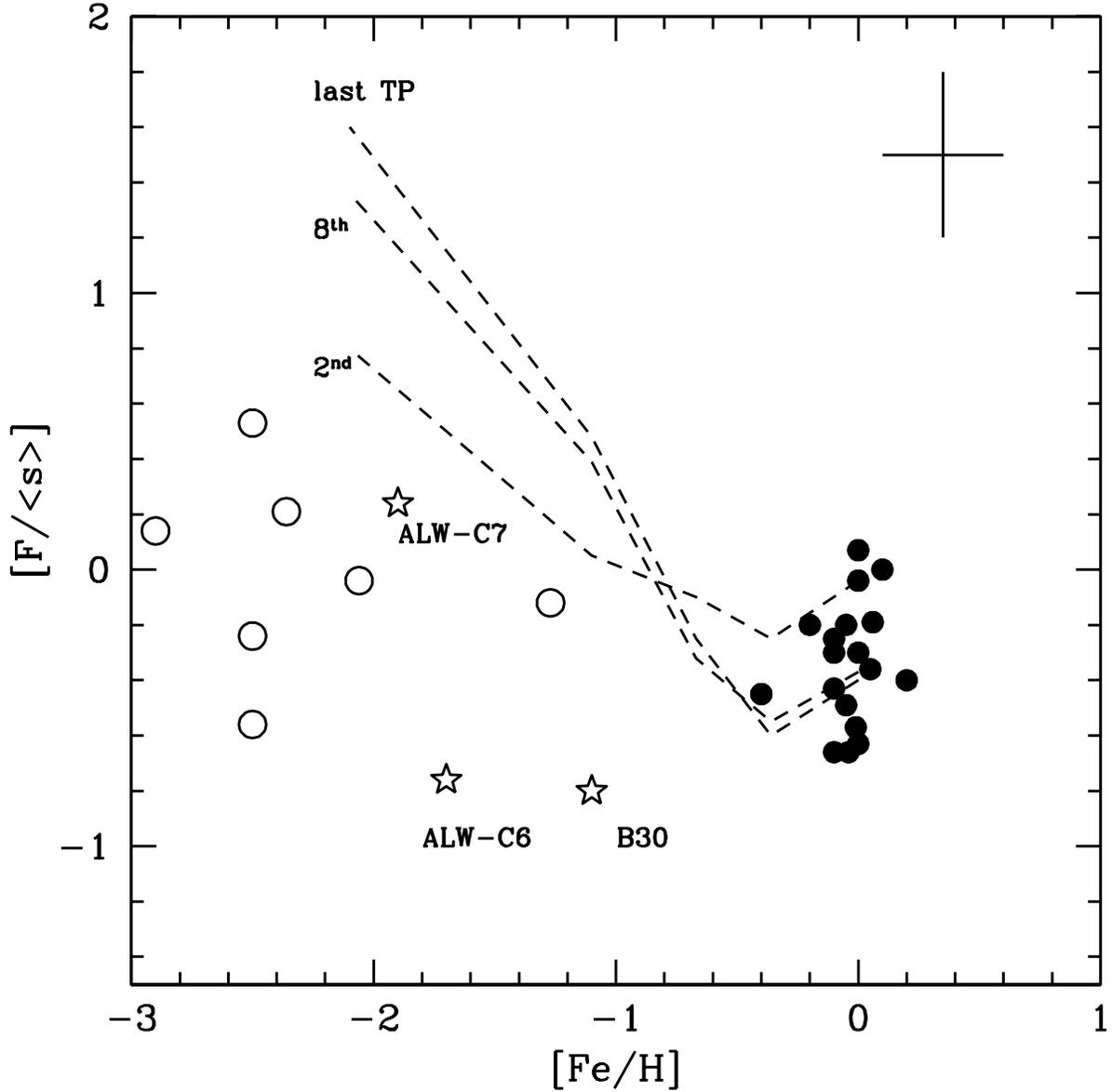}
\caption{Observed [F/$<$s$>$] vs. [Fe/H] ratios in intrinsic AGB carbon stars (symbols as in 
Figure 2) and extrinsic CEMP-s stars (open circles, from Lucatello et al. 2011 and
references therein) compared with theoretical predictions (dashed lines) at several
thermal pulses ($2^{nd}, 8^{th}$ and last TP) for a 1.5 M$_\odot$ TP-AGB model with 
different metallicities. \label{fig4}}
\end{figure}

\begin{deluxetable}{lccccccccccccc}
\tabletypesize{\scriptsize}
\rotate
\tablewidth{0pt}
\tablecaption{Data and Abundances for Program Stars\tablenotemark{a}}
\tablehead{
\colhead{Star}           & \colhead{$K$}  &
\colhead{Exp. time (s)} &  S/N                &
\colhead{T$_{\rm eff}$} & [Fe/H]            & 
\colhead{C/N\tablenotemark{b}/O}       &  \colhead{log $\epsilon($C$-$O)}    & 
\colhead{$^{16}$O/$^{17}$O}  & \colhead{log $\epsilon$(F)} & 
\colhead{log $\epsilon$(Na)} &  \colhead{[F/Fe]}&
\colhead{[Na/Fe]}& [$<$s$>$/Fe]\tablenotemark{d}}
\startdata
LMC TRM88  &10.30&5400 & 55 &3400   &-0.6   &8.20/7.25/8.05&  7.63 & 425   & 4.85  & 5.40  & +0.85& -0.24&\nodata\\
LMC MSX663\tablenotemark{c}&10.20& 5400& 57 &\nodata&\nodata&\nodata/\nodata/\nodata&\nodata&\nodata&\nodata&\nodata&\nodata&\nodata&\nodata\\
SMC BMB-B30&10.70&6300 & 65 & 3000  & -1.1  &7.70/6.83/7.60&  7.05 & 440   & 3.60  &5.20   & +0.14& 0.06&+0.90\\
SMC GM780\tablenotemark{d}&10.25&3600&67&2000&-1.3&7.45/6.50/7.40&6.81&\nodata&$<$3.26 &$<$4.80 &$<$0.0&$<$-0.14&\nodata\\
Carina ALW-C6&12.30&9000&35 & 3400  &-1.7   &7.60/6.10/6.70&  7.54 &\nodata&3.70   &$<$4.50&+0.84&$<$0.0&+1.60\\
Carina ALW-C7& 12.70&7200&30 & 3200  &-1.9   &7.40/5.83/6.40&  7.35 &\nodata&4.40   &5.30   &+1.74&+0.96&+1.50\\

\enddata
\tablenotetext{a}{The adopted solar abundances are from Asplund et al. (2009).}
\tablenotetext{b}{N abundances are scaled to the stellar metallicity.}
\tablenotetext{c}{The average s-element enhancements aren taken from de Laverny et al. (2006) for BMB B30, 
and Abia et al. (2008) for the stars in Carina.}
\tablenotetext{d}{The star LMC MSX663 shows no spectral lines in the 2.3350 $\mu$m region and was
discarded for abundance analysis (see text).}
\tablenotetext{e}{The amostpheric parameters adopted for this star and the derived abundances 
have to be considered as very uncertain (see text).}
\end{deluxetable}

\end{document}